\documentclass[manuscript]{aastex6}

\newcommand{\wind}{\textit{Wind}}
\newcommand{\stereoa}{\textit{STEREO A}}
\newcommand{\soho}{\textit{SOHO${/}$}LASCO}
\newcommand{\kms}{km s$^{-1}$}
\newcommand{\rsun}{$R_\sun$}
\usepackage{amsmath}
\shorttitle{An unexpected geomagnetic storm produced by a stealth CME}
\shortauthors{He et al.}

\begin{document}

\title{A Stealth CME bracketed between slow and fast wind producing unexpected geo-effectiveness}

\author{Wen He\altaffilmark{1,2}, Ying D. Liu\altaffilmark{1,2},
        Huidong Hu\altaffilmark{1,2}, Rui Wang\altaffilmark{1,3}, 
        and Xiaowei Zhao\altaffilmark{1,2}}
\altaffiltext{1}{State Key Laboratory of Space Weather,
        National Space Science Center,
        Chinese Academy of Sciences, Beijing 100190, China;
        \href{mailto:liuxying@spaceweather.ac.cn}{liuxying@swl.ac.cn}}
\altaffiltext{2}{University of Chinese Academy of Sciences, Beijing 100049, China}
\altaffiltext{3}{W.W. Hansen Experimental Physics Laboratory, Stanford University, Stanford, CA, USA}

\begin{abstract}

We investigate how a weak coronal mass ejection (CME) launched on 2016 October 8 without obvious signatures in the low corona produced a relatively intense geomagnetic storm. Remote sensing observations from \textit{SDO}, \textit{STEREO}{} and \textit{SOHO} and in situ measurements from \wind{} are employed to track the CME from the Sun to the Earth. Using a graduated cylindrical shell (GCS) model, we estimate the propagation direction and the morphology of the CME near the Sun. CME kinematics are determined from the wide-angle imaging observations of \stereoa{} and are used to predict the CME arrival time and speed at the Earth. We compare ENLIL MHD simulation results with in situ measurements to illustrate the background solar wind where the CME was propagating. We also apply a Grad--Shafranov technique to reconstruct the flux rope structure from in situ measurements in order to understand the geo-effectiveness associated with the CME magnetic field structure. Key results are obtained concerning how a weak CME can generate a relatively intense geomagnetic storm: (1) there were coronal holes at low latitudes, which could produce high speed streams (HSSs) to interact with the CME in interplanetary space; (2) the CME was bracketed between a slow wind ahead and a HSS behind, which enhanced the southward magnetic field inside the CME and gave rise to the unexpected geomagnetic storm. 
\end{abstract}

\keywords{(Sun:) solar wind -- solar--terrestrial relations -- Sun: coronal mass ejections (CMEs)}

\section{Introduction} \label{sec:intro}

Coronal mass ejections (CMEs) are spectacular eruptions of plasma and magnetic fields from the solar corona. Interplanetary CMEs (ICMEs) are the heliospheric counterpart of CMEs and are often associated with prolonged southward magnetic field components. A southward interplanetary magnetic field is closely related to geomagnetic activities \citep{1991Gosling}. Because of the strong connection between the evolution of a CME in interplanetary space and its geo-effectiveness at the Earth (\citealt{2017Kilpua}, and references therein), understanding CME evolution in the inner heliosphere is of critical importance for space weather forecasting. 

Slow CMEs (with speeds below the typical solar wind speed, i.e., 400 \kms) are generally thought to be not geo-effective. However, the interplanetary evolution of CMEs including slow ones can be complicated, such as interactions with other CMEs and the highly structured solar wind (e.g., \citealt{Gopalswamy2001b, 2012Lugaz, 2012Liu, 2014Liu, 2017Liu, 2015Kataoka}). As \citet{2015Liu} envision, a combination of different solar wind structures can result in enhanced geo-effectiveness. Indeed, slow CMEs spend a long time travelling in the Sun-Earth space, so they have a high potential to interact with other solar wind structures, providing critical means to enhance their geo-effectiveness \citep{2016Liu}. \citet{2004Tsurutani} find that some slow ICMEs surprisingly caused intense geomagnetic storms. However, it is still unclear how slow CMEs lead to enhanced geo-effectiveness by interacting with other solar wind structures. 

Slow CMEs can occur as ``stealth'' events. Generally, CMEs are observed as bright transient features in a white-light coronagraph and are associated with a variety of phenomena in the low corona, such as solar flares, filaments eruptions, EUV waves, and coronal dimmings. However, some CMEs do not have low-coronal signatures (LCSs) to determine their source regions. With the help of stereoscopic observations from Solar TErestrial RElations Observatory (STEREO; \citealt{2008Kaiser}) which is comprised of two spacecraft (\stereoa{} and B), we are able to eliminate the ambiguity of the frontside or backside origin of halo CMEs. \citet{2009Robbrecht} present a detailed analysis of a frontside CME without obvious LCSs in EUV and H$\alpha$ based on \textit{STEREO}{} observations. They suggest that the CME was a streamer-blowout CME and originated from a high altitude of the corona, so it did not leave clear on-disk signatures. Following this event, the term ``stealth CME'' has been used to refer to a class of CMEs without LCSs on the disk. Subsequent studies confirm that stealth CMEs tend to be slow and occur close to coronal holes \citep{2010Ma, 2014Huy}. As a kind of slow CMEs, stealth CMEs have a high probability to interact with other solar wind structures like high speed streams (HSSs) and may cause unexpected geo-effectiveness \citep{2014Kilpua, 2016Liu}. 

Solar cycle 24 is one of the weakest solar cycles ever recorded. Reduced solar activity is expected to lead to decreased geomagnetic activity. However, slow CMEs are frequently observed (e.g., see the Richardson and Cane ICME list at \url{http://www.srl.caltech.edu/ACE/ASC/DATA/level3/icmetable2.htm}). Another characteristic of solar cycle 24 is that coronal holes, especially those near the solar equator, also occur frequently. Therefore, the interaction between slow CMEs (including stealth events) and HSSs from coronal holes is expected to be a frequent phenomenon in solar cycle 24. Enhanced geo-effectiveness may thus arise. In this paper, we present a comprehensive analysis of a geo-effective stealth CME that occurred on 2016 October 8, combining remote sensing and in situ observations from \textit{SDO}, \textit{STEREO}, \textit{SOHO}, and \wind. A relatively intense geomagnetic storm with the minimum $D_{st}$ index of --104 nT occurred despite a weak CME. This geomagnetic storm is unexpected. We track the evolution of this stealth CME from the Sun to the Earth and pay particular attention to how such a slow CME caused an intense geomagnetic storm. We examine the coronal signatures in Section \ref{1}, the interplanetary propagation characteristics in Section \ref{2}, and in situ measurements and associated geo-effectiveness in Section \ref{3}. The results are summarized and discussed in Section \ref{4}. This work provides insights on the Sun-to-Earth evolution of CMEs and how enhanced geo-effectiveness can arise from the interplanetary evolution.

\section{CME near the Sun} \label{1}

\begin{figure}
\epsscale{1.1}
\plottwo{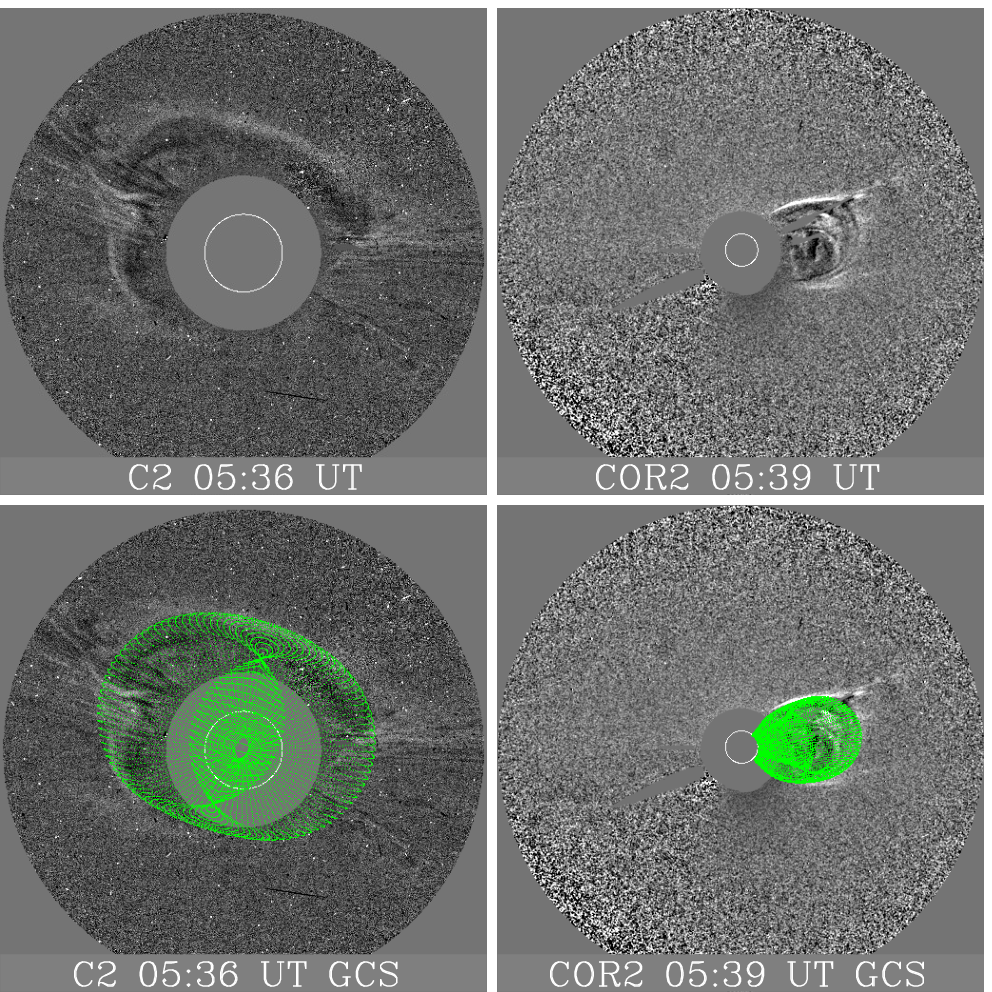}{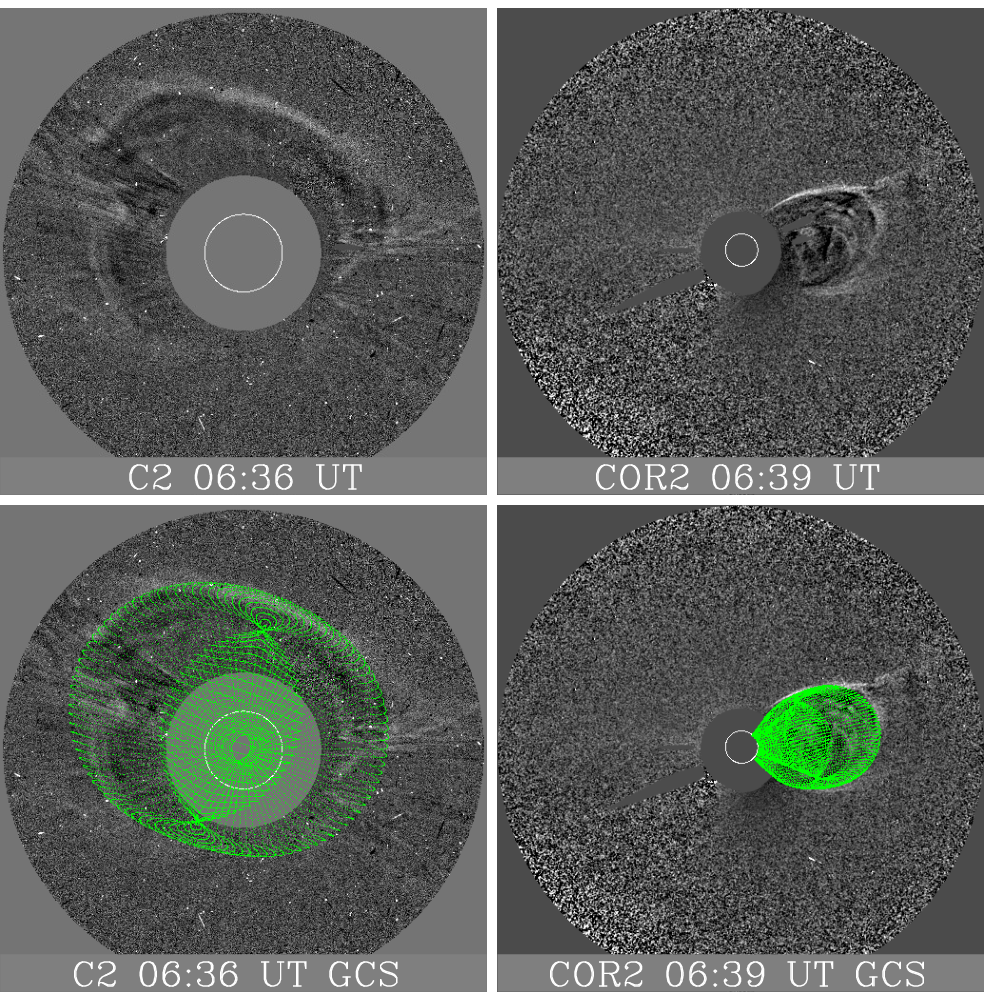}
\caption{\label{f1}Running difference coronagraph images and corresponding GCS modeling results (green grids) from \textit{LASCO} C2 and \stereoa/COR2 on 2016 October 9. } 
\end{figure}

\begin{figure}
\epsscale{0.8}
\plotone{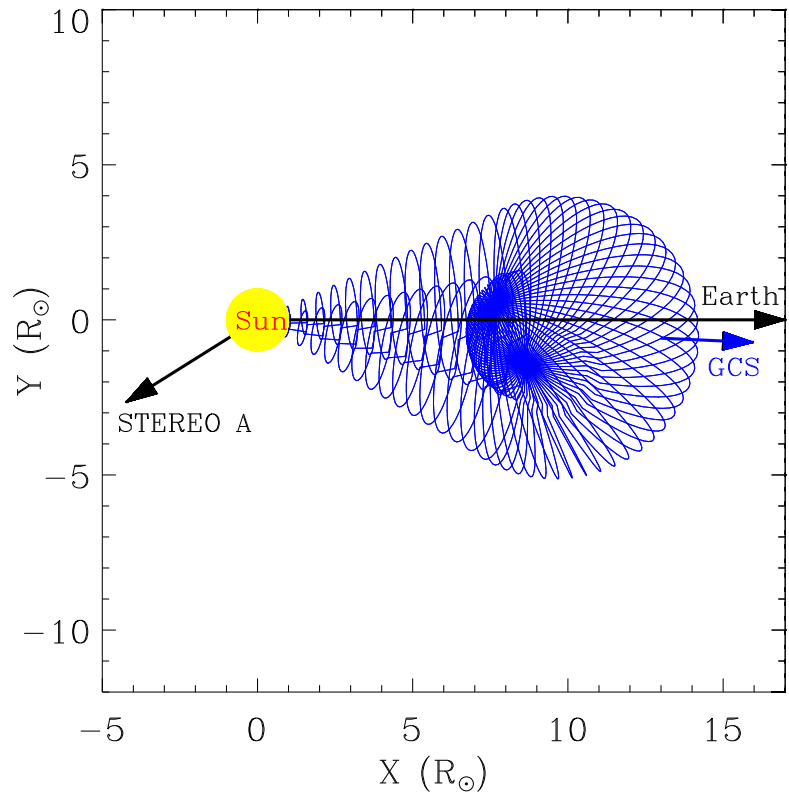}
\caption{\label{f2}Geometry of the CME flux rope (blue line) projected onto the ecliptic plane on 2016 October 8. The blue arrow indicates the propagation direction of the CME derived from the GCS model. The leading edge of the CME is at 14.5 \rsun{} from the sun. The directions of the Earth and \stereoa{} are marked by the black arrows.}
\end{figure}

An interplanetary shock arrived the Earth at 21:15 UT on 2016 October 12, and a G2 geomagnetic storm ($D_{st}$ $<$ --100 nT) occurred. Tracing back to the Sun, we find a CME observed by \stereoa{} and \textit{SOHO} simultaneously during October 8 to 9. The coronagraph images are shown in Figure \ref{f1}. This is the most probable CME that is associated with the in situ signatures at the Earth (also see details below). It was first seen in the northeastern sector of \soho{} and evolved into a halo CME on October 9. \stereoa{} was 142.9\degr{} east of the Sun-Earth line at a distance of 0.96 au from the Sun, and caught a sided view of this CME in the western part of COR2. We use a graduated cylindrical shell (GCS) model \citep{2006Thernisien} to simulate the CME with simultaneous observations from \textit{SOHO}{} and \stereoa. The GCS model can determine the direction of propagation, flux rope orientation and height based on the coronagraph images. It has been successfully applied to \soho{} and \textit{STEREO}/SECCHI observations (e.g., \citealt{2009Thernisien, 2010bLiu, 2017Liu, 2017Hu, 2017Zhao}). The GCS model fits the CME observations from the two vantage points very well (see lower panels of Figure \ref{f1}). The modeled CME structure is projected onto the ecliptic plane, as shown in Figure \ref{f2}. The propagation direction of the CME obtained from the GCS model does not change much within 14.5 \rsun, which is on average about 3\degr{} east of the Sun-Earth-line and about 11\degr{} north of the ecliptic plane. The tilt angle of the CME flux rope is small (about 18\degr), so the flux-rope axis roughly lies in the ecliptic plane. If the flux rope does not rotate in interplanetary space, a spacecraft near the Earth would see a change in the direction of the meridional magnetic field component across the flux rope. The speed of the CME leading edge is about 360 \kms{} obtained from a linear fit of the GCS distances.

\begin{figure}
\epsscale{0.8}
\plotone{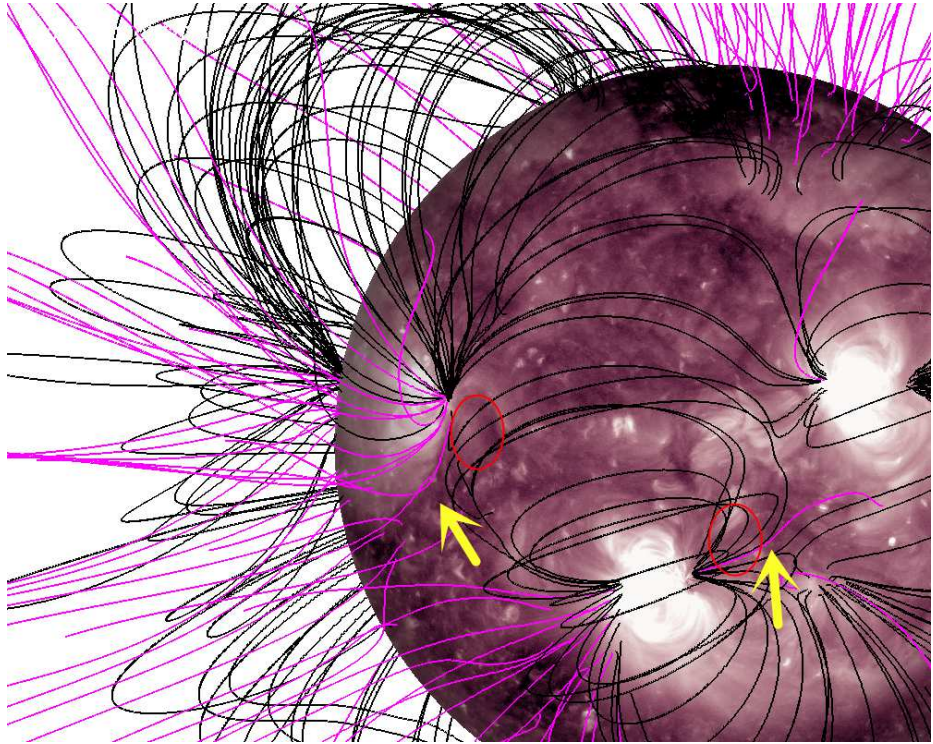}
\caption{\label{f3}PFSS extrapolation of global coronal magnetic fields mapped onto the AIA 211 \AA{} image at 22:36 UT on 2016 October 8. The magenta and black lines represent the open and closed magnetic field lines, respectively. The yellow arrows indicate the locations of two coronal holes near the solar equator. Two red circles denote the two dimming regions proposed by \citet{2017Nitta}. The source region of the CME, however, is unclear.}
\end{figure}

There were no obvious signatures in the low corona, so this event is probably a ``stealth CME''. Utilizing a potential field source surface (PFSS) method based on full-Sun magnetograms \citep{1969Schatten, 2003Schrijver}, we derive the global coronal magnetic field topology and identify the locations of coronal holes with open magnetic field lines. Figure \ref{f3} shows the result of PFSS extrapolation mapped onto a 211 \AA{} image from the Atmosphere Imaging Assembly (AIA; \citealt{2012Lemen}) on \textit{SDO}. The open magnetic field lines together with the AIA image suggest that there are mainly two coronal holes near the solar equator (indicated by yellow arrows). \citet{2017Nitta} try to search for the CME source region and find two possible dimming regions from long-time (14-hr interval) percentage difference AIA 211 \AA{} images. The locations of the two dimming regions are marked as red circles in Figure \ref{f3}. The east dimming region is close to the eastern coronal hole extending to the solar equator. A coronal hole may change the propagation direction of a CME in the corona \citep{2009Gopalswamy}. The HSS from the coronal hole may influence the propagation of the CME in interplanetary space as well (e.g., \citealt{2016Liu}). Since the source region of the stealth CME cannot be determined unambiguously, we are unable to quantify the possible influence of the coronal holes on the CME propagation. 

\section{Evolution in Interplanetary Space} \label{2}

\subsection{CME Kinematics}
\begin{figure}[htb] 
\epsscale{0.8}
\centerline{\includegraphics[scale=0.75]{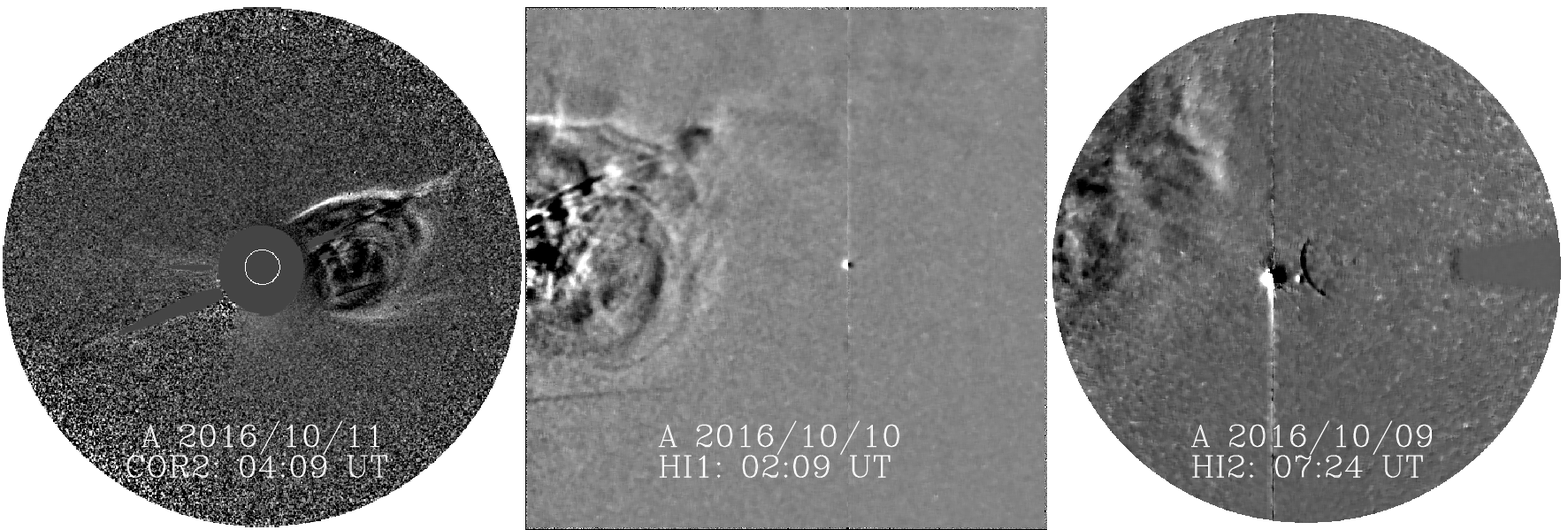}}
\centerline{\includegraphics[scale=0.75]{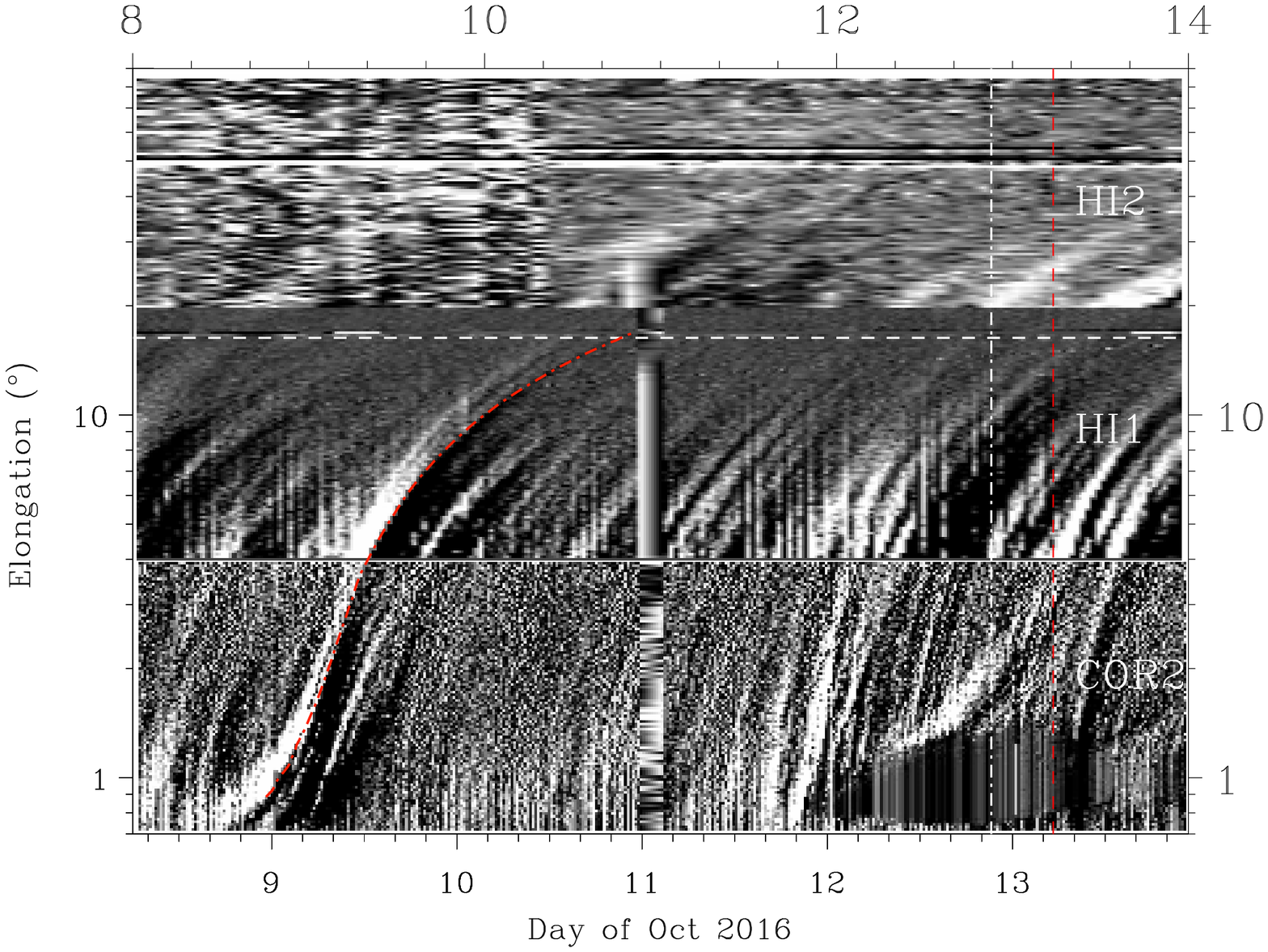}}
\caption{\label{f4}Top: Evolution of the 2016 October 8 CME. From left to right, the panels show the running difference images of COR2, HI1 and HI2 of \stereoa{} at different times. Bottom: Time-elongation map constructed from running difference images of COR2, HI1 and HI2 on \stereoa{} along the ecliptic plane. The red curve indicates the track of the CME, along which the elongation angles are extracted. The horizontal white line marks the elongation angle of the Earth. The white vertical line denotes the observed arrival time of the shock at the Earth, while the red vertical line represents the estimated arrival time from HM approximation.}
\end{figure}

Figure \ref{f4} (top) illustrates the synoptic view of the CME in interplanetary space from COR2, HI1 and HI2 of \stereoa. The angular field of view is 0\degr.7--4\degr{} around the Sun for COR2, 20\degr{} square centered at 14\degr{} from the center of the Sun for HI1, and 70\degr{} centered at 53\degr7 for HI2 \citep{2008Howard}. The CME appears as a slow, streamer-blowout event in COR2 and can be tracked to large distances from the Sun. By stacking the running-difference images of COR2, HI1, and HI2 within a slit along the ecliptic plane, we obtain the time-elongation map \citep{2008Sheeley, 2009Davies, 2010aLiu} as shown in Figure \ref{f4} (bottom). Elongation angles of the CME in the ecliptic plane are extracted from the track in the time-elongation map. There are several methods that can be used to convert elongation angles to radial distances from the Sun (see a summary in \citealt{2010bLiu}). \citet{2009Lugaz} propose a harmonic mean (HM) approximation, which assumes a spherical structure attached to the Sun for a CME and what is seen by a spacecraft is the segment tangent to the line of sight. We derive the CME kinematics with the HM approximation using the propagation angle (3\degr{} east of the Sun-Earth line) from GCS modeling. An explicit assumption is that the propagation angle does not change in interplanetary space. The CME kinematics can also be derived from an F$\beta$ approximation (assuming a compact CME structure), which, however, yields an unphysical late acceleration because of the large observation angle of \stereoa{} (see \citealt{2013Liu}).

The CME kinematics from HM approximation within 100 \rsun{} are shown in Figure \ref{f5}. The speed profile shows a gradual acceleration up to about 20 \rsun{} and thereafter a roughly constant value at about 400 \kms. This is a typical speed profile for slow CMEs as discussed by \citet{2016Liu}. The CME is probably accelerated by the forward drag force of the ambient solar wind and finally reaches the typical speed of the ambient solar wind. Note that in this case we can only track the CME to about 90 \rsun. It is not clear if the speed would have a significant change beyond 90 \rsun{} due to interactions with a fast solar wind stream from behind (see below). The interaction with a fast stream from behind, however, may not change the speed significantly, since what we are tracking is the leading edge of the CME. Given the roughly constant speed after about 20 \rsun, we use a linear extrapolation of the distances after 20 \rsun{} to predict the CME arrival time at the Earth. The predicted arrival time is about 05:00 UT on October 13, which is about 8 hr later than the observed shock arrival time at the Earth. The final speed of $\sim$400 \kms{} is also consistent with the average speed in the sheath region of the ICME (about 437 \kms). These results unambiguously connect the CME on October 8--9 as seen by \soho{} and \textit{STEREO} to the ICME detected during October 12--14 at L1 and to the subsequent geomagnetic storm.

\subsection{WSA-ENLIL Simulations}
\begin{figure}
\plottwo{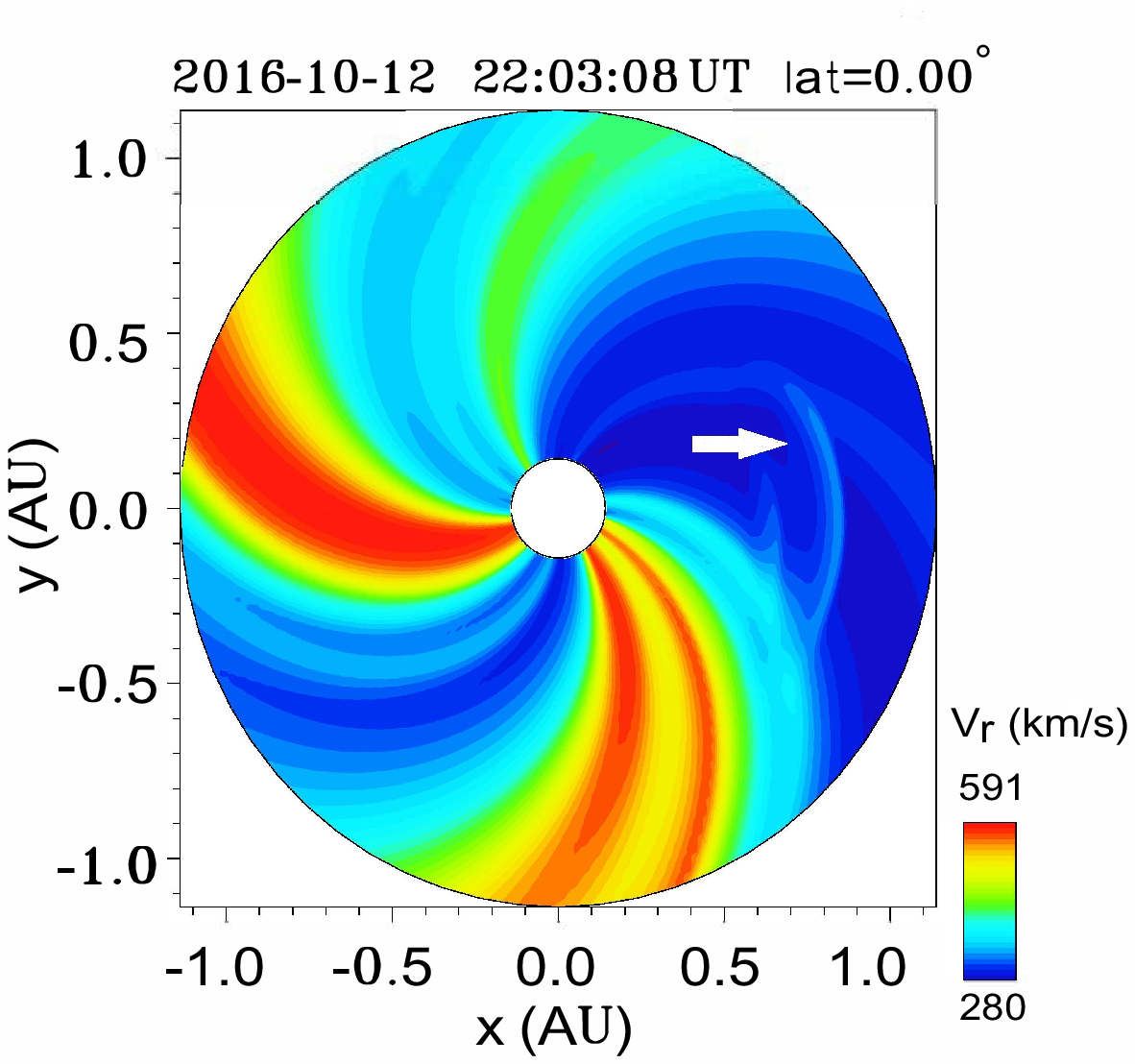}{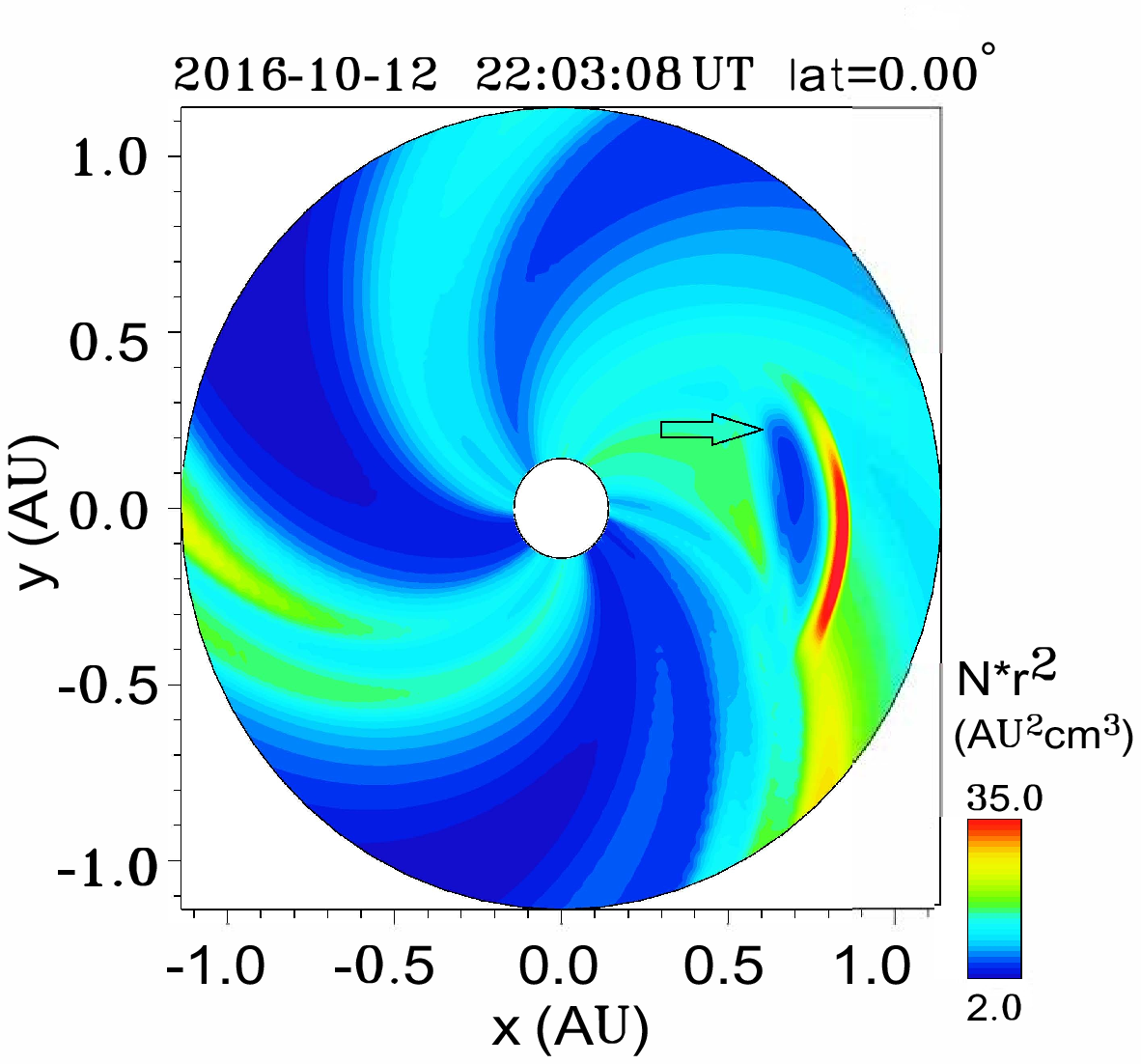}
\caption{\label{f6}Speed (left) and density (right) distributions from WSA-ENLIL+Cone model in the ecliptic plane. The values of the speed and density are scaled by the color bar at the right corner of each panel. The arrow in each panel indicates the location of the simulated ICME. The simulation run used in the study is performed at the NASA/CCMC and is available upon request at \url{https://ccmc.gsfc.nasa.gov/database_SH/Wen_He_071417_SH_1.php}. }
\end{figure}

In order to determine the distribution of the ambient solar wind and the environment where the CME was propagating, we request WSA-ENLIL MHD modeling from the Community Coordinated Modeling Center (CCMC; \url{https://ccmc.gsfc.nasa.gov/}). The WSA-ENLIL global 3D MHD model provides a time-dependent description of the background solar wind plasma and magnetic field as well as an inserted CME. We use the GCS results as input for the simulation, including the time when the CME crosses the inner boundary at 21.5 \rsun, CME half angular width, radial speed and propagation direction. Figure \ref{f6} shows the simulation results out to 1.15 au at 22:03 UT on October 12. Obviously, an Earth-directed CME is propagating into a slow solar wind and is followed by a HSS ($\sim$500\kms) \footnote{The ENLIL simulation underestimates the speed of the HSS, which is about 700 \kms{} as observed at 1 au.}. By examining the speed distribution in different longitudes and latitudes, we find that the HSS probably originates from the eastern coronal hole in Figure \ref{f3}. The CME would be compressed since it is bracketed between a slow and fast wind. Indeed, the density distribution (see Figure \ref{f6} right) suggests that the CME is inside a compression region. This compression would lead to an enhanced magnetic field inside the CME observed at 1 au, although the simulation itself does not contain a flux rope \citep{2004Odstrcil}.

\section{ICME Properties near the Earth} \label{3}

\begin{figure}
\plotone{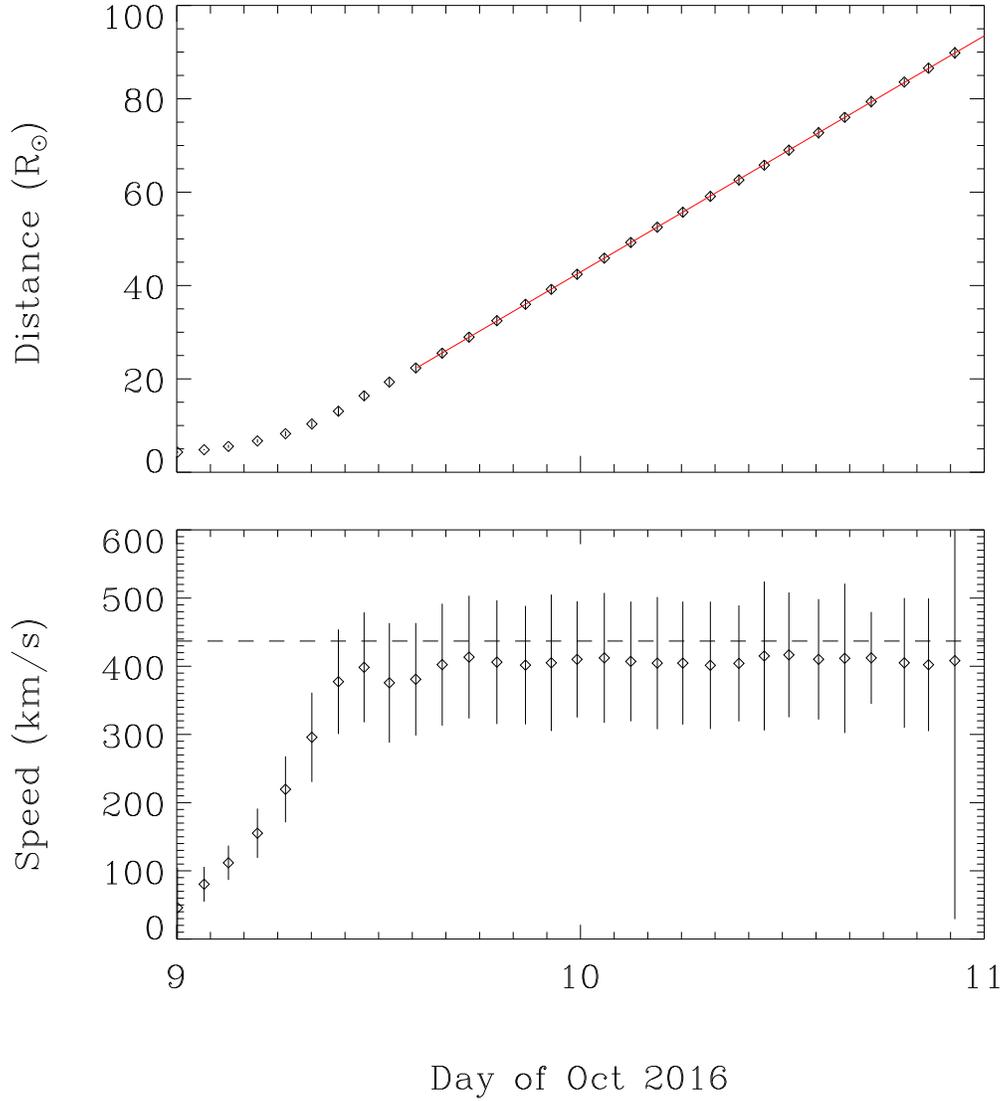}
\caption{\label{f5}Radial distance and speed profiles of the CME derived from HM approximation. The red line denotes a linear fit of the distances after 20 \rsun. The speeds are calculated from the distances using a numerical differentiation method. The horizontal line in the bottom panel marks the average speed of 437 \kms{} in the sheath region of the ICME observed at 1 au. }
\end{figure}

\begin{figure}
\epsscale{0.8}
\plotone{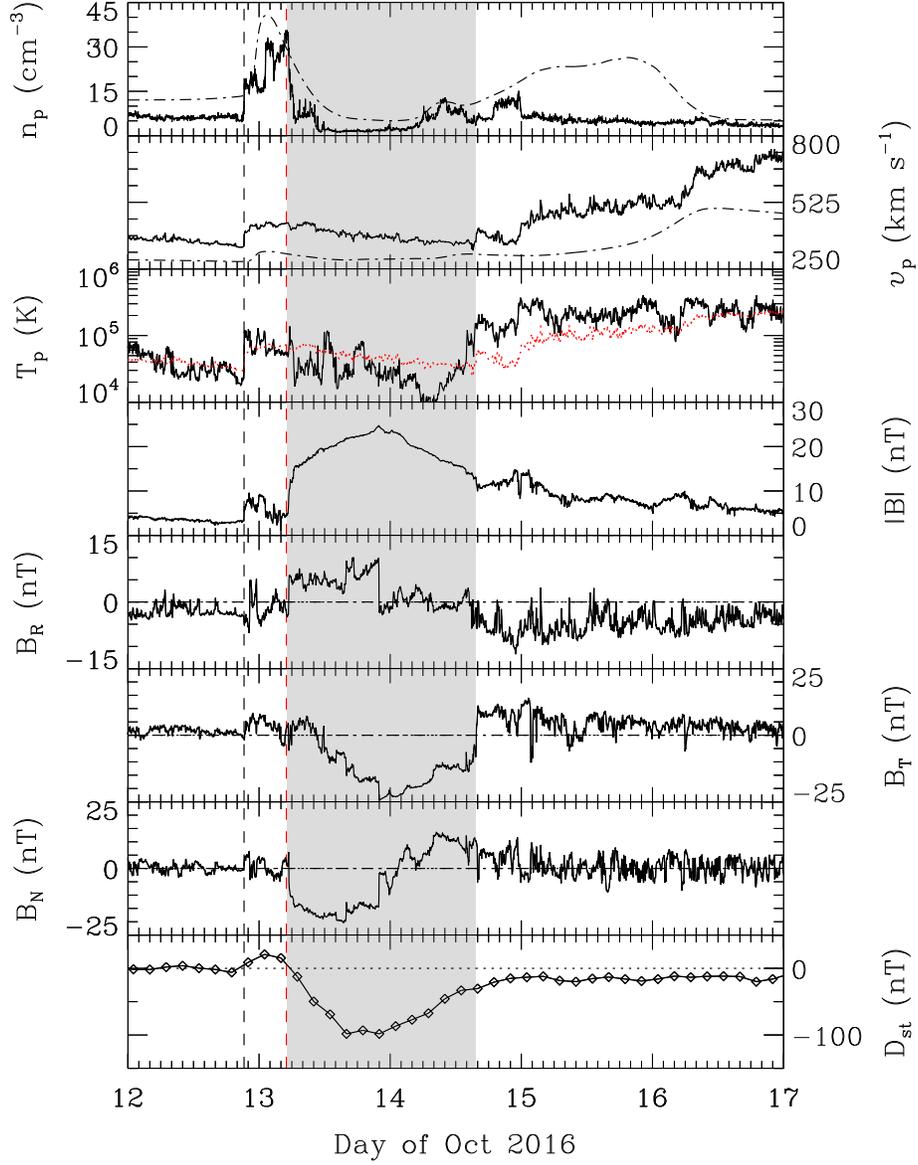}
\caption{\label{f7}Solar wind parameters associated with the 2016 October 8 CME observed at \wind. From top to bottom, the panels show the proton density, bulk speed, proton temperature, magnetic field strength and components, and $D_{st}$ index, respectively. The bulk speed and density from the ENLIL simulation at the Earth are overplotted after being shifted 15 hr earlier. The red dotted line in the third panel denotes the expected proton temperature calculated from the observed speed \citep{1987Lopez}. The shaded region indicates the flux rope interval determined by the GS reconstruction. The black vertical dashed line marks the observed arrival time of the shock, while the red vertical dashed line is the predicted arrival time from measured CME kinematics. }
\end{figure}

Figure \ref{f7} shows the in situ measurements at \wind. A shock passed \wind{} around 21:15 UT on October 12, which is $\sim$8 hr earlier than the predicted arrival time from the measured CME kinematics (Figure \ref{f5}). The shock was likely produced by the expansion of the CME into an even slower solar wind environment. Typical signatures of magnetic cloud are observed, such as depressed proton temperatures, enhanced magnetic fields, and long-lasting smooth rotation of the field components. The average speed in the sheath between the shock and ejecta is about 437 \kms, comparable to the predicted speed (Figure \ref{f5}). The shock arrival time predicted by the WSA-ENLIL model is about 15 hr late compared with the observed shock arrival time. After shifting the simulation results by about 15 hr, we can see that the speed and density profiles from the simulation generally agree with the in situ measurements overall, although the ENLIL simulation underestimates the speed and overestimates the density for this event. In the ENLIL simulation, there is a gradual speed increase after the magnetic cloud before the fast stream arrives, which accords with the in situ observations. The velocity measurements indicate that the CME is sandwiched between the slow and fast wind, which is also consistent with the ENLIL simulation result (Figure \ref{f6}). The magnetic field profile of the magnetic cloud appears symmetric with a peak near the center. This symmetric profile is likely due to the sandwiching between the slow and fast streams, which prevents the magnetic cloud from expanding. Within the ICME, the magnetic field strength is as high as 25 nT while the southward field component reaches --21 nT. In general, strong ejecta magnetic fields tend to be associated with fast CMEs (e.g., \citealt{2010Richardson}). For a slow CME, such a magnetic field is unusually strong. Again, the magnetic field inside the ICME must have have been enhanced because the CME is inside a compression region between a slow and fast wind. This can explain the unexpected geo-effectiveness (with the minimum $D_{st}$ index of --104 nT) despite a weak stealth CME. 

\begin{figure}
\plotone{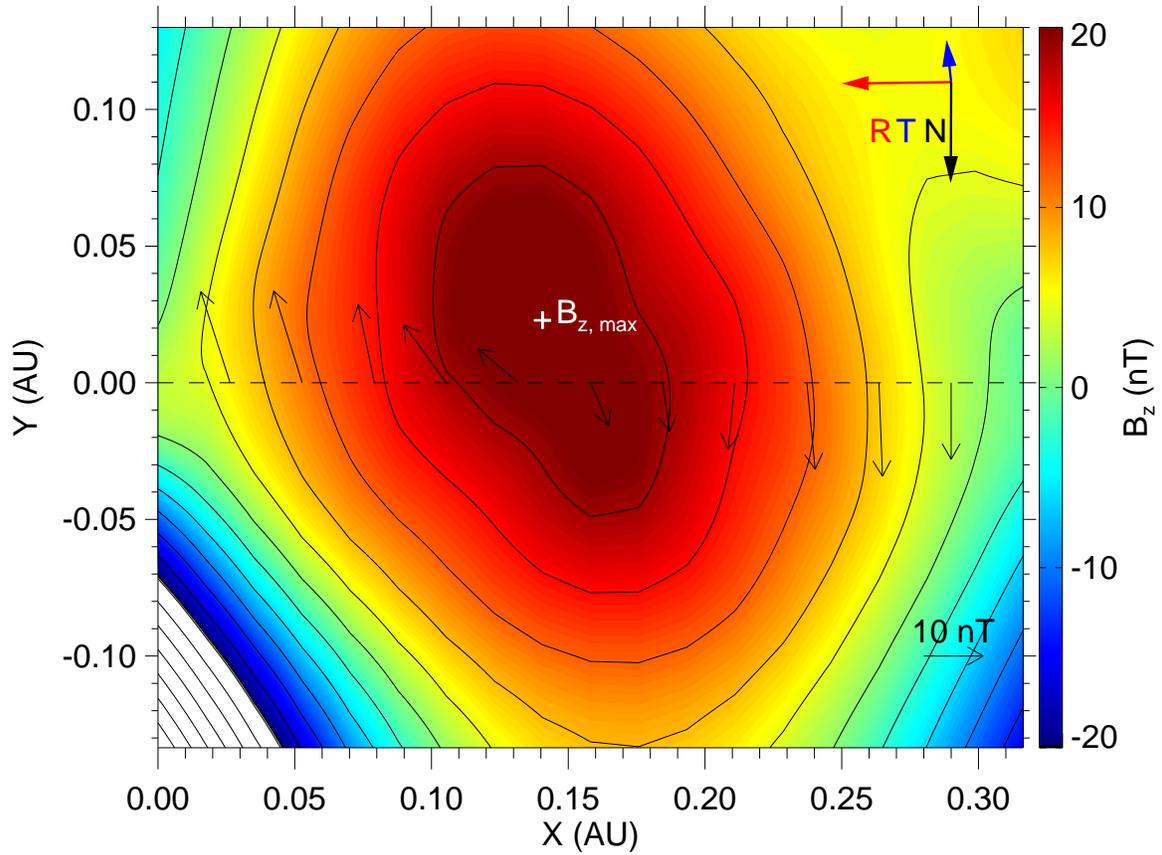}
\caption{\label{f8}Reconstructed cross section of the flux rope at \wind. Black contours show the distribution of the vector potential, and the color shading indicates the value of the axial magnetic field. The location of the maximum axial field is labeled by the white cross. The horizontal dashed line marks the trajectory of \wind. The thin black arrows denote the direction and magnitude of the observed magnetic fields projected onto the cross section. The thick colored arrows display the projected RTN directions.}
\end{figure}

Enhanced and prolonged southward magnetic fields associated with ICMEs are important triggers of geomagnetic storms, and depend on the flux rope orientation as well as the axial and azimuthal components of the flux rope magnetic field \citep{2015Liu}. We use a Grad--Shafranov (GS) technique \citep{1999Hau, 2002Hu}, which has been validated by multi-spacecraft measurements  \citep{2008Liu, 2009Mostl}, to reconstruct the flux rope structure at the Earth. Figure \ref{f8} gives the reconstruction results, which show how the magnetic field features control the geomagnetic storm activity. This is a left-handed flux-rope with the axis elevation angle of about --23\degr{} and the azimuthal angle of about 272\degr{} in RTN coordinates. The flux-rope tilt angle is similar to that determined from GCS modeling near the Sun. The maximum azimuthal field component is comparable to the maximum axial component. From these results, the geomagnetic storm was mainly caused by the azimuthal magnetic field component of the flux rope in combination with a slightly southward flux-rope orientation.

\section{summary and discussions}\label{4}
   
We have performed a comprehensive analysis of the 2016 October 8 stealth CME and how it produced unexpected geo-effectiveness, covering remote sensing observations from \textit{SDO}, \textit{STEREO}{} and \textit{SOHO}{} and in situ measurements at \wind. A PFSS method together with EUV observations has been used to examine the coronal environment of the CME. The Sun-to-Earth evolution of the CME is analyzed using both observations and modeling. Finally, a GS reconstruction method is employed to understand the ICME flux-rope structure and how it controls the geomagnetic activity. Key findings are obtained concerning how a stealth CME evolved in interplanetary space and how the evolution resulted in an unexpected geomagnetic storm. Below, we summarize the results and discuss their implications. 

The CME was a stealth event without clear low coronal signatures. It shows a gradual acceleration followed by a nearly constant speed around the average solar wind level in interplanetary space. This is a speed profile of a typical slow CME \citep{2016Liu}. All these suggest that the event may not be geo-effective. However, the CME became geo-effective at 1 au and led to a G2 magnetic storm with $D_{st}$ minimum of --104 nT. Through the reconstruction of the flux rope  structure near the Earth, we find that the geomagnetic storm was mainly caused by the azimuthal component of the flux-rope magnetic field in combination with a slightly southward flux-rope orientation (also see \citealt{2005Huttunen}) .

EUV observations and PFSS extrapolation of the coronal magnetic field suggest that there were low-latitude coronal holes, which could produce high-speed streams to interact with the CME in interplanetary space. Indeed, both ENLIL MHD simulations and in situ measurements at the Earth indicate that the CME was inside a compression region bracketed by a slow and fast wind. This explains the unusually strong magnetic field inside the ICME observed at 1 au and the unexpected geo-effectiveness. These results, again, suggest that slow CMEs have a high potential to interact with other solar wind structures in the Sun-Earth space due to their slow motion \citep{2016Liu}. The results also indicate the crucial importance of CME interplanetary evolution for accurate space weather prediction. This seems particularly important for solar cycle 24, although weak, because slow CMEs and low-latitude coronal holes are ubiquitous features of the solar cycle.  A weak solar cycle does not necessarily implies weak geomagnetic activity (also see \citealt{2015bKilpua}). 

\acknowledgments

The research was supported by NSFC under grants 41774179, 41374173 and 41604146, and the Specialized Research Fund for State Key Laboratories of China. We acknowledge the use of data from {\textit{STEREO}, \textit{SDO}, \textit{SOHO}, \wind, and \textit{GONG}}. The MHD simulation results are provided by the Community Coordinated Modeling Center at Goddard Space Flight Center through their public Runs on Request system (\url{http://ccmc.gsfc.nasa.gov}). The ENLIL Model was developed by D. Odstr$\check{c}$il at the University of Colorado at Boulder.
\clearpage

\clearpage

\clearpage

\clearpage

\clearpage

\clearpage

\clearpage

\clearpage


\clearpage
\bibliographystyle{aasjournal}
\bibliography{article}

\begin{thebibliography}{}
\expandafter\ifx\csname natexlab\endcsname\relax\def\natexlab#1{#1}\fi





\bibitem[Davies et al.(2009)]{2009Davies} Davies, J.~A., Harrison, R.~A., Rouillard, A.~P., et al.\ 2009, \grl, 36, L02102 

\bibitem[D'Huys et al.(2014)]{2014Huy} D'Huys, E., Seaton, D.~B., Poedts, S., \& Berghmans, D.\ 2014, \apj, 795, 49 

\bibitem[Gopalswamy et al.(2001)]{Gopalswamy2001b} Gopalswamy, N., Yashiro, S., Kaiser, M.~L., Howard, R.~A., \& Bougeret, J.-L.\ 2001, \apjl, 548, L91 

\bibitem[Gopalswamy et al.(2009)]{2009Gopalswamy} Gopalswamy, N., M{\"a}kel{\"a}, P., Xie, H., Akiyama, S., \& Yashiro, S.\ 2009, \jgr, 114, A00A22 



\bibitem[Gosling et al.(1991)]{1991Gosling} Gosling, J.~T., McComas, D.~J., Phillips, J.~L., \& Bame, S.~J.\ 1991, \jgr, 96, 7831 


\bibitem[Hau \& Sonnerup(1999)]{1999Hau} {Hau}, L.-N., \& {Sonnerup}, B.~U.~{\"O}.\ 1999, \jgr, 104, 6899 

\bibitem[Howard et al.(2008)]{2008Howard} Howard, R.~A., Moses, J.~D., Vourlidas, A., et al.\ 2008, \ssr, 136, 67 


\bibitem[Hu et al.(2017)]{2017Hu} Hu, H., Liu, Y.~D., Wang, R., et al.\ 2017, \apj, 840, 76 

\bibitem[Hu \& Sonnerup(2002)]{2002Hu} Hu, Q., \& Sonnerup, B.~U.~{\"O}.\ 2002, \jgr, 107, 1142 

\bibitem[Huttunen, et al.(2005)]{2005Huttunen} Huttunen, K.~E.~J., Schwenn, R., Bothmer, V., et al.\ 2005, Annales Geophysicae, 23, 625.

\bibitem[Kaiser et al.(2008)]{2008Kaiser} Kaiser, M.~L., Kucera, T.~A., Davila, J.~M., et al.\ 2008, \ssr, 136, 5 

\bibitem[Kataoka et al.(2015)]{2015Kataoka} Kataoka, R., Shiota, D., Kilpua, E., \& Keika, K.\ 2015, \grl, 42, 5155 

\bibitem[Kilpua et al.(2014)]{2014Kilpua} Kilpua, E.~K.~J., Mierla, M., Zhukov, A.~N., et al.\ 2014, \solphys, 289, 3773 

\bibitem[Kilpua et al.(2015)]{2015bKilpua} Kilpua, E.~K.~J., Olspert, N., Grigorievskiy, A., et al.\ 2015, \apj, 806, 272 

\bibitem[Kilpua et al.(2017)]{2017Kilpua} Kilpua, E.~K.~J., Balogh, A., von Steiger, R., \& Liu, Y.~D.\ 2017, \ssr, 212, 1271 

\bibitem[Lemen et al.(2012)]{2012Lemen} Lemen, J.~R., Title, A.~M., Akin, D.~J., et al.\ 2012, \solphys, 275, 17 

\bibitem[Liu et al.(2008)]{2008Liu} Liu, Y., Luhmann, J.~G., Huttunen, K.~E.~J., et al.\ 2008, \apjl, 677, L133 

\bibitem[Liu et al.(2010a)]{2010aLiu} Liu, Y., Davies, J.~A., Luhmann, J.~G., et al.\ 2010a, \apjl, 710, L82 

\bibitem[Liu et al.(2010b)]{2010bLiu} Liu, Y., Thernisien, A., Luhmann, J.~G., et al.\ 2010b, \apj, 722, 1762 

\bibitem[Liu et al.(2012)]{2012Liu} Liu, Y.~D., Luhmann, J.~G., M{\"o}stl, C., et al.\ 2012, \apjl, 746, L15 

\bibitem[Liu et al.(2013)]{2013Liu} Liu, Y.~D., Luhmann, J.~G., Lugaz, N., et al.\ 2013, \apj, 769, 45 

\bibitem[Liu et al.(2014)]{2014Liu} Liu, Y.~D., Luhmann, J.~G., Kajdi{\v c}, P., et al.\ 2014, Nature Communications, 5, 3481 

\bibitem[Liu et al.(2015)]{2015Liu} Liu, Y.~D., Hu, H., Wang, R., et al.\ 2015, \apjl, 809, L34 

\bibitem[Liu et al.(2016)]{2016Liu} Liu, Y.~D., Hu, H., Wang, C., et al.\ 2016, \apjs, 222, 23 

\bibitem[Liu et al.(2017)]{2017Liu} Liu, Y.~D., Zhao, X., \& Zhu, B.\ 2017, \apj, 849, 112 

\bibitem[Lopez(1987)]{1987Lopez} Lopez, R.~E.\ 1987, \jgr, 92, 11189 

\bibitem[Lugaz et al.(2009)]{2009Lugaz} Lugaz, N., Vourlidas, A., \& Roussev, I.~I.\ 2009, Annales Geophysicae, 27, 3479 

\bibitem[Lugaz et al.(2012)]{2012Lugaz} Lugaz, N., Farrugia, C.~J., Davies, J.~A., et al.\ 2012, \apj, 759, 68 

\bibitem[Ma et al.(2010)]{2010Ma} Ma, S., Attrill, G.~D.~R., Golub, L., \& Lin, J.\ 2010, \apj, 722, 289 

\bibitem[M{\"o}stl et al.(2009)]{2009Mostl} M{\"o}stl, C., Farrugia, C.~J., Miklenic, C., et al.\ 2009, \jgr, 114, A04102 

\bibitem[Nitta \& Mulligan(2017)]{2017Nitta} {Nitta}, N.~V., \& {Mulligan}, T.\ 2017, \solphys, 292, 125 

\bibitem[Odstrcil et al.(2004)]{2004Odstrcil} Odstrcil, D., Riley, P., \& Zhao, X.~P.\ 2004, \jgr, 109, A02116 

\bibitem[Richardson \& Cane(2010)]{2010Richardson} Richardson, I.~G. \& Cane, H.~V.\ 2010, \solphys, 264, 189.

\bibitem[Robbrecht et al.(2009)Robbrecht, Patsourakos, and Vourlidas]{2009Robbrecht} {Robbrecht}, E., {Patsourakos}, S., \& {Vourlidas}, A.\ 2009, \apj, 701, 283 


\bibitem[Sheeley et al.(2008)]{2008Sheeley} Sheeley, N.~R., Jr., Herbst, A.~D., Palatchi, C.~A., et al.\ 2008, \apj, 675, 853-862 

\bibitem[Schatten et al.(1969)]{1969Schatten} Schatten, K.~H., Wilcox, J.~M., \& Ness, N.~F.\ 1969, \solphys, 6, 442 

\bibitem[Schrijver \& De Rosa(2003)]{2003Schrijver} {Schrijver}, C.~J., \& {De Rosa}, M.~L.\ 2003, \solphys, 212, 165 
\bibitem[Tsurutani, et al.(2004)]{2004Tsurutani} Tsurutani, B.~T., Gonzalez, W.~D., Zhou, X.-Y., et al.\ 2004, Journal of Atmospheric and Solar-Terrestrial Physics, 66, 147.

\bibitem[Thernisien et al.(2006)]{2006Thernisien} Thernisien, A.~F.~R., Howard, R.~A., \& Vourlidas, A.\ 2006, \apj, 652, 763 

\bibitem[Thernisien et al.(2009)]{2009Thernisien} Thernisien, A., Vourlidas, A., \& Howard, R.~A.\ 2009, \solphys, 256, 111 
 





\bibitem[Zhao et al.(2017)]{2017Zhao} Zhao, X., Liu, Y.~D., Hu, H., \& Wang, R.\ 2017, \apj, 837, 4 

\end{thebibliography}

\end{document}